\documentclass[english,conference,draftcls,onecolumn]{IEEEtran}
\usepackage{draftwatermark}
\SetWatermarkText{DRAFT}

\usepackage[utf8x]{inputenc}
\usepackage{amsmath}
\usepackage{amssymb}
\usepackage[numbers]{natbib}
\usepackage{array}
\usepackage{subfig}
\usepackage{url}

\usepackage[english,british]{babel}
\usepackage[keeplastbox]{flushend}

\usepackage[colorinlistoftodos]{todonotes}

\begin{document}

\title{Towards Anticipation of Architectural Smells using Link Prediction Techniques}

\author{
 \IEEEauthorblockN{J. Andr\'{e}s D\'{i}az-Pace\IEEEauthorrefmark{1}, Antonela Tommasel\IEEEauthorrefmark{2}, Daniela Godoy\IEEEauthorrefmark{3}}
 \IEEEauthorblockA{ISISTAN Research Institute, CONICET-UNICEN University. Argentina\\
\IEEEauthorrefmark{1}\emph{andres.diazpace@isistan.unicen.edu.ar},
\IEEEauthorrefmark{2}\emph{antonela.tommasel@isistan.unicen.edu.ar},
\IEEEauthorrefmark{3}\emph{daniela.godoy@isistan.unicen.edu.ar}
}
}

\maketitle

\begin{abstract}
Software systems naturally evolve, and this evolution often brings design problems that cause system degradation. Architectural smells are typical symptoms of such problems, and several of these smells are related to undesired dependencies among modules. The early detection of these smells is important for developers, because they can plan ahead for maintenance or refactoring efforts, thus preventing system degradation. Existing tools for identifying architectural smells can detect the smells once they exist in the source code. This means that their undesired dependencies are already created. In this work, we explore a forward-looking approach that is able to infer groups of likely module dependencies that can anticipate architectural smells in a future system version. Our approach considers the current module structure as a network, along with information from previous versions, and applies link prediction techniques (from the field of social network analysis). In particular, we focus on dependency-related smells, such as Cyclic Dependency and Hub-like Dependency, which fit well with the link prediction model. An initial evaluation with two open-source projects shows that, under certain considerations, the predictions of our approach are satisfactory. Furthermore, the approach can be extended to other types of dependency-based smells or metrics. 

\end{abstract}


\section{Introduction}
\label{sec:intro} 

Software systems naturally evolve due to changes in their requirements
and operating environment. Along this evolution, the amount
and complexity of the interactions among the software elements of a system
are likely to increase, with a consequent effect on the system design structure, which tends to erode over time \cite{Hochstein:2005:CAD:1709687.1709719,DBLP:journals/jss/SilvaB12}. For instance, modules might become dependent on each other because of a new functionality being implemented. Design erosion symptoms are often related to high coupling and undesired dependencies in the module structure. The so-called architectural smells \cite{conf/qosa/GarciaPEM09} are (poor) design decisions that contribute to system erosion. Several of these smells, such as cycles \cite{Melton2007}, involve particular patterns of undesired dependencies. The early detection of such smells is important for developers, so that
they can plan ahead for maintenance or refactoring efforts, and
preserve the quality and integrity of the system.

In this context, several tools exist for helping developers to manage system dependencies and also detect some types of smells, including: LattixDSM, SonarQube, Sonargraph, HotspotDetector, or Arcan, to name a few  \cite{Hochstein:2005:CAD:1709687.1709719,DBLP:conf/icsa/FontanaPRTZN17,Mo2015HotspotPT}. These tools normally extract dependency graphs from source code and might compute metrics, which altogether serve to identify candidate architectural smells in the code. However, a limitation of this kind of tools is that they are only able to detect the smells once their underlying dependencies actually exist in the source code. Although identifying smells is important, we argue that being able to anticipate them (before they appear) is more helpful, because developers can take actions early to prevent the smells. Furthermore, once introduced in the code, violations to design rules (like those caused by smells) can be difficult to fix by developers \cite{Hochstein:2005:CAD:1709687.1709719}. 

In this article, we propose a proactive approach that leverages on link prediction (LP) techniques for inferring likely configurations of architectural smells. In prior work~\cite{diazpace_icsa_2018}, we investigated whether LP techniques can rely on structural features from previous system versions to predict module dependencies for the next version, by means of Machine Learning (ML) techniques. That work was concerned with individual dependencies only. Here, we further develop those ideas by looking at groups of dependencies (both actual and predicted ones), which enable us to spot architectural smells from the literature \cite{conf/qosa/GarciaPEM09,DBLP:conf/icsm/FontanaPRZ16}. These smells are considered at the package level. The approach makes two contributions: i) it defines strategies for identifying specific types of smells based on the outputs of a ML classification model (which only predicts individual dependencies); and ii) it incorporates content-based features in the  classification model for boosting its predictions. Thus, these contributions improve the capabilities of our prediction approach.

We provide an instantiation of the approach for two well-known architectural smells: Cyclic Dependency and Hub-like Dependency \cite{DBLP:journals/ibmrd/Marinescu12,DBLP:journals/sigsoft/Tracz15a}, and evaluate its performance in two open-source Java systems. We consider different system versions for training the classification model and testing the quality of the predictions (both dependencies and smells). The results so far show that the approach is able to identify most smell instances (i.e., good recall), at the expenses of also identifying some mistaken smells to be analyzed (and eventually discarded) by a developer (i.e., affecting precision).

The rest of the article is organized into 7 sections. Section 2 gives background information about architectural smells, and motivates the prediction of dependencies for smell configurations. Section 3 discusses how the prediction of package dependencies is cast as a link prediction problem, and further, to a classification problem. We also explain the notion of topological and content-based features. Section 4 presents the main building blocks of our prediction approach. Section 5 describes the performed experimental study with two Java open-source systems. The main results and lessons learned of the study are discussed in Section 6. Section 7 covers related work. Finally, Section 8 presents the conclusions and outlines future work. 

\section{From Dependencies to Architectural Smells}
\label{sec:background}

Software systems often exhibit design problems, which can be either introduced during development or along their evolution. These problems, also known as \textit{architectural smells} \cite{conf/qosa/GarciaPEM09}, have a negative impact on the quality of the system, as they degrade its design structure. A smell usually comes from a poorly-understood or sub-optimal design decision. Different architectural smells have been catalogued in the literature \cite{conf/qosa/GarciaPEM09, DBLP:journals/sigsoft/Tracz15a,DBLP:journals/ibmrd/Marinescu12,Lippert2006}. Of particular relevance to this work are the so-called \textit{dependency-based smells} \cite{conf/qosa/GarciaPEM09}, which involve interactions among system components. These smells occur when one or more components violate design principles or rules, and often manifest themselves as \textit{undesired dependencies} in the source code \cite{Hochstein:2005:CAD:1709687.1709719}. Two examples of such smells are: \textit{Cyclic Dependency} (CD) \cite{DBLP:journals/ibmrd/Marinescu12} and \textit{Hub-like Dependency} (HLD) \cite{DBLP:journals/sigsoft/Tracz15a}. CD and HLD smells can be identified at the class or at the package level \cite{DBLP:conf/icsa/FontanaPRTZN17}. In this work, we focus on their characteristics and detection strategies for packages.

\subsubsection*{Cyclic Dependency} 
\begin{figure}
\centering \includegraphics[width=0.78\columnwidth]{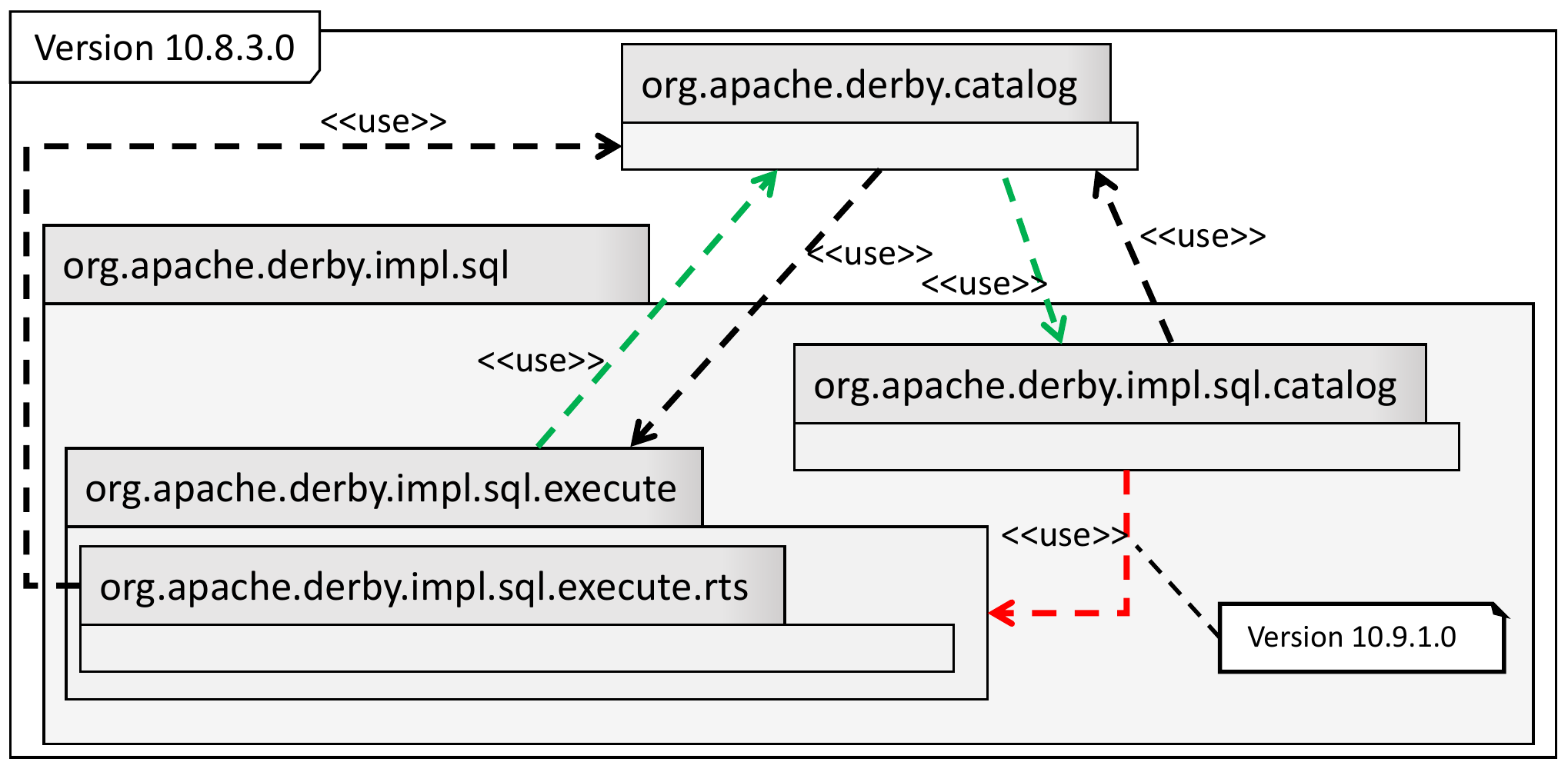}
\caption{Example of Cyclic Dependency (Apache Derby)}\label{fig:example-cyclic-dependency}
\end{figure}

In this smell, various components directly or indirectly depend on each other to function properly. For example, Figure~\ref{fig:example-cyclic-dependency} depicts a cycle among three packages of the Apache Derby project (excerpt). The packages are connected by means of usage relations (dependencies). The cycle (denoted by green and red arrows) did not exist in version 10.8.3.0 but was introduced in version 10.9.1.0 due to the addition of a dependency between \texttt{org.apache.derby.impl.sql.catalog} and \texttt{org.apache.derby.impl.sql.execute.rts} (in red in the figure). This relation is a case of an undesired dependency, from the perspective of the CD smell. In general, the chain of relations among packages breaks the desirable acyclic nature of a subsystem's dependency structure. Thus, the components involved in a cycle can be hard to maintain, test or reuse in isolation. Cycles might have different shapes \cite{Melton2007, DBLP:conf/aswec/Al-MutawaDMM14}, and some cycles might be more harmful for the system health than others. The strategies for detecting cycles in the package structure are based on the DFS graph algorithm \cite{DBLP:conf/icsa/FontanaPRTZN17}. 

\subsubsection*{Hub-like Dependency}\label{sec:hld} 
\begin{figure}
\centering \includegraphics[width=0.98\columnwidth]{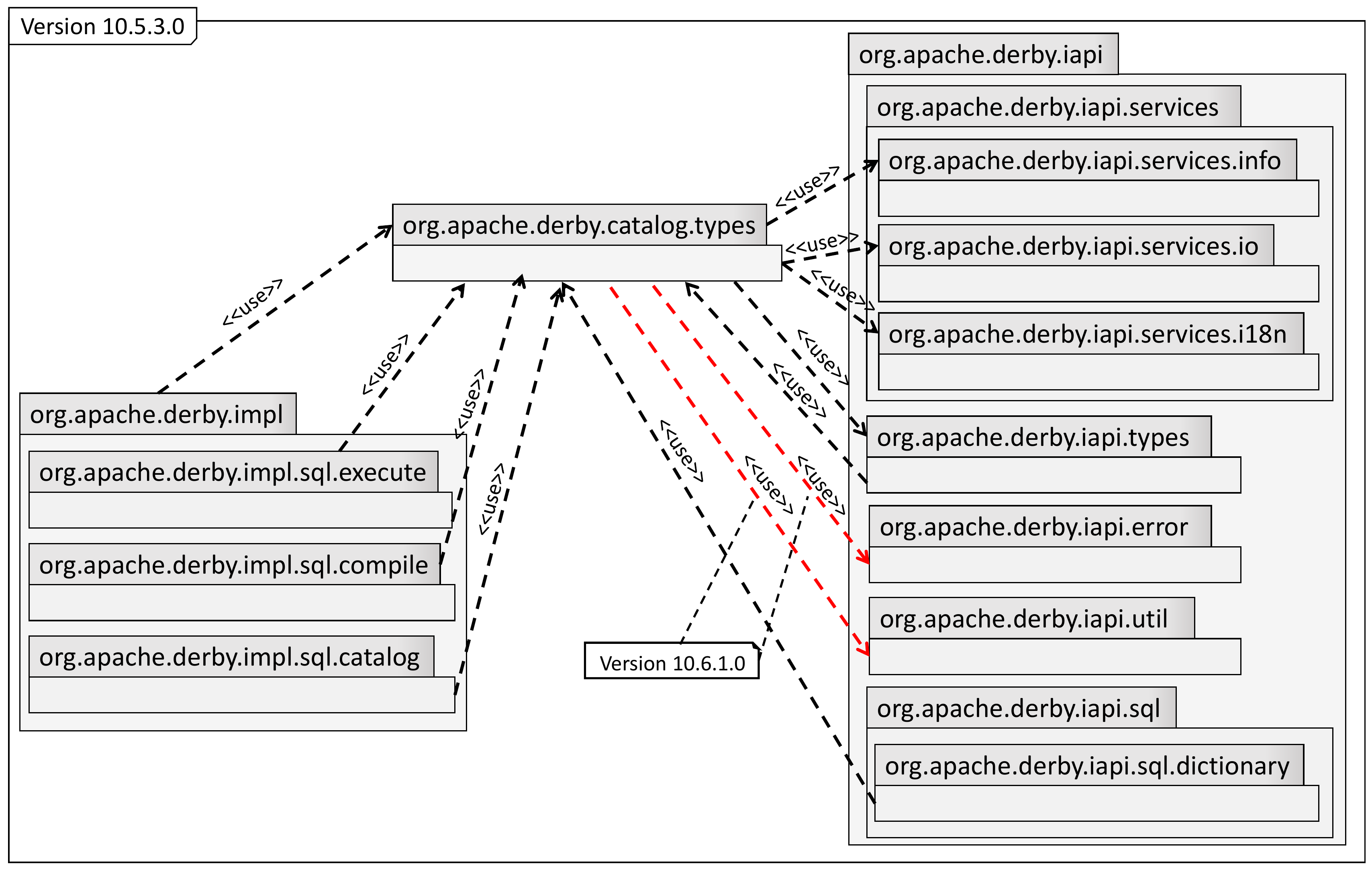}
\caption{Example of Hub-like Dependency (Apache Derby)}\label{fig:example-hub-dependency}
\end{figure}

This smell arises when a component has outgoing and ingoing dependencies with a large number of other components. For example, Figure~\ref{fig:example-hub-dependency} shows that package \texttt{org.apache.derby.catalog.types} uses 4 packages (right side) and is used by 6 other packages in version 10.5.3.0 of Apache Derby (excerpt). Then, in version 10.6.1.0, the central package requires two additional outgoing dependencies (in red in the figure), which altogether transform the package into a hub. These two relations are undesired dependencies, from the perspective of the HLD smell. The strategy for detecting potential hubs in a package network follows from \cite{DBLP:journals/sigsoft/Tracz15a}. It first computes the median of the number of incoming and outgoing dependencies of all packages. Then, for each package, it checks if both its incoming and outgoing dependencies are greater than the incoming and outgoing medians respectively, and finally checks whether the difference between the incoming and outgoing dependencies is less than a fraction of the total dependencies of that package. According to this detection strategy, \texttt{org.apache.derby.catalog.types} at version 10.5.3.0 is not a hub, but when the package structure changes in version  10.6.1.0 (because of the two dependencies added), then the detection strategy marks \texttt{org.apache.derby.catalog.types} as a hub. 

In the two examples above, the dependencies judged as "undesirable" are due to new functionality allocated to existing classes or new classes being created under existing packages. These dependencies inadvertently appear (in the source code) from one system version to another. We assume here an architecture compliance process that periodically checks whether the current version of the system implementation satisfies a set of design rules, such as avoiding architectural smells, and reports any issues to developers. 

Particular dependency-based smells can be detected (and also refactored) by means of several tools, such as: HotspotDetector  \cite{Mo2015HotspotPT}, Arcan  \cite{DBLP:conf/icsa/FontanaPRTZN17}, Sonargraph, or Structure101, among others. Although useful, these tools are mostly reactive, in the sense that they are intended for scenarios in which the smells are realized in the code. Along this line, removing a smell is not always straightforward for developers, due to the additional efforts for refactoring undesired dependencies while ensuring that the system continues to function properly. Our work is motivated by the vision of a proactive tool able to spot patterns of dependencies in the code, or "quasi-smells", in which a (future) insertion of a few dependencies would precipitate the smells. If the most likely dependencies can be predicted, developers could benefit from early "warnings" on potential architectural smells in the system.

\section{Link Prediction Techniques}
\label{sec:link-prediction}
A key observation about the appearance of dependencies contributing
to architectural smells, like CD and HLD, is that, despite changes
occurring at the class level, the package structure (or module level)
remains more or less stable across system versions while dependencies
among packages keep being added. There are, of course, a few exceptions
such as: the initial versions of the system, in which the main structure
and functionality is fleshed out, or when a refactoring takes place
in a given version. Overall, by analyzing the package structure and
its evolution over time, it is possible to predict which dependencies
are likely to appear between pairs of packages. In this context, we
resort to \emph{link prediction} (LP) techniques from the field
of social network analysis (SNA).

LP adapts SNA techniques for studying to what extent the evolution
of a network can be modeled by using its intrinsic \emph{features} ~\cite{Liben-Nowell:2007:LPS:1241540.1241551}. This involves
inferring "missing" links between pairs
of nodes in a network (traditionally represented as a graph) based
on the observable interactions (or links) among nodes and node attributes
~\cite{Liben-Nowell:2007:LPS:1241540.1241551}. An assumption here
is that a software system can be seen as a network of software elements
(e.g., modules, classes, methods) that behaves similarly to a social network,
at a given level of abstraction (e.g., at the package level in Java
systems).

A prerequisite for applying LP is to transform the system under analysis into a dependency graph. More formally, a dependency graph is a graph $DG\left(V,E\right)$, where each node $v\in V$ represents a module,
and each edge (or link) $e\left(v,v'\right)\in E$ represents a dependency
from node $v$ to $v'$ ($v,v'\in V$). Since we deal with Java systems,
nodes correspond to packages while edges represent relations
between those packages. For this work, we limit dependencies to usage relations and assume that the dependency graph is extracted
from the Java source code via static analysis techniques. For simplicity, we also consider each package as an individual module, although in practice some low-level packages are intended for code organization purposes and might not  define module boundaries. More formally,
our LP task takes a $DG\left(V,E\right)$ at time $n$, and then infers
the edges that will be added to $DG\left(V,E\right)$ at time $n+1$.
Let $U$ be the set of all possible edges among nodes in $DG\left(V,E\right)$.
The LP task generates a list $R$ of all possible edges in $U-E$,
and indicates whether each edge (in $R$) is present in $DG\left(V,E\right)$
at time $n+1$.

LP in social networks is based on the principle of homophily~\cite{doi:10.1146/annurev.soc.27.1.415},
which states that interactions between similar individuals occur at
a higher rate than those among dissimilar ones. In our context, this
would mean that similar packages (according to some criteria) have
a higher chance to establish dependencies than dissimilar packages.
Most techniques for the LP problem are based on graph \emph{topological}
features that assess similarity between pairs of nodes~\cite{Liben-Nowell:2007:LPS:1241540.1241551}. Nonetheless, other types of features are also possible. In the following, we discuss topological and content-based features.

\subsubsection*{Topological features} 
\begin{figure}
\begin{centering}
\includegraphics[width=0.95\columnwidth]{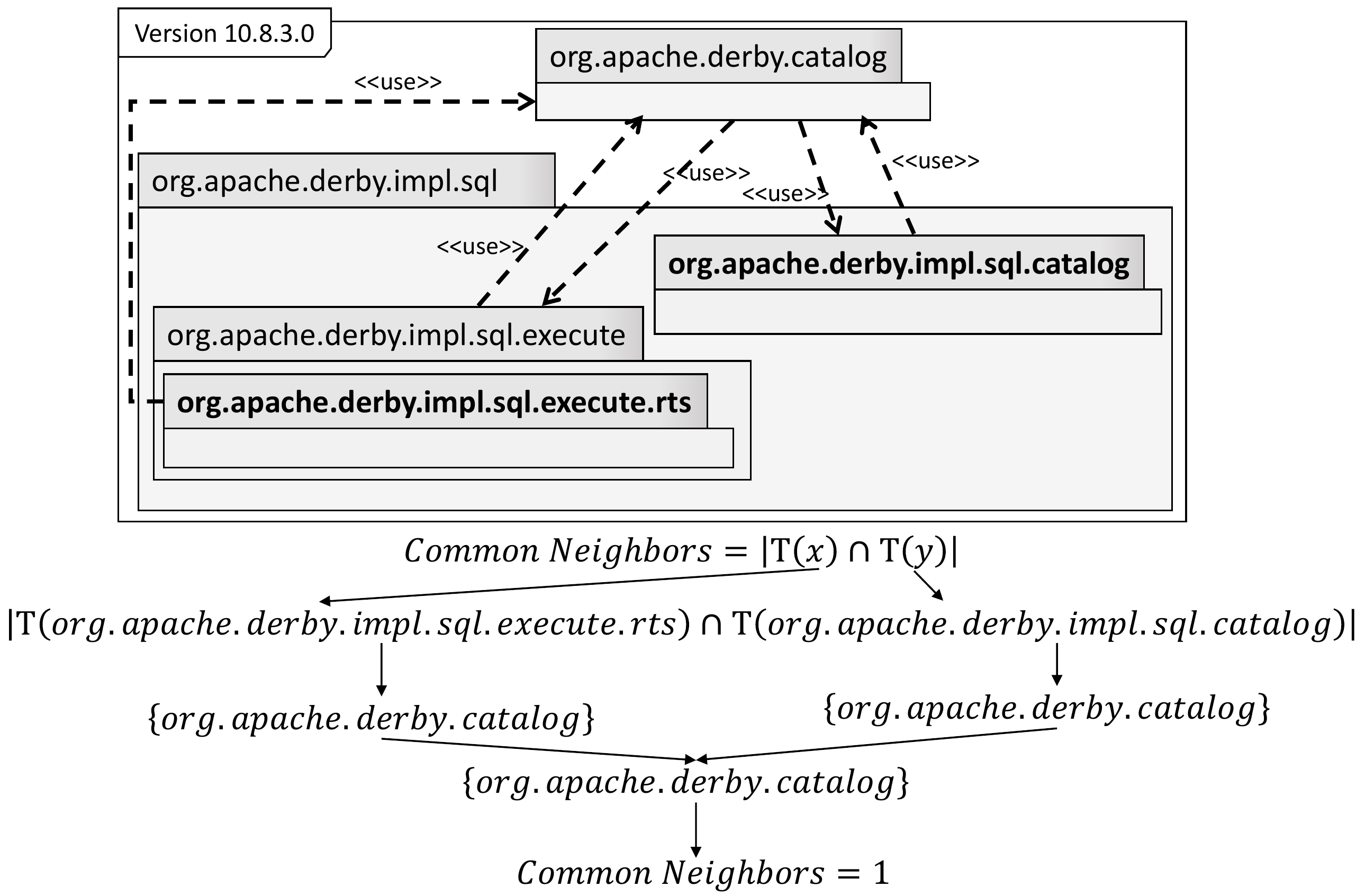}
\par\end{centering}

\caption{\label{fig:Example-of-Topological}Computing a Topological Feature
for two Packages}
\end{figure}

These features are related to the graph structure and the role that the nodes (and their edges) play in that structure~\cite{Liben-Nowell:2007:LPS:1241540.1241551}. For instance, \emph{Common Neighbors} is defined as the number of common adjacent nodes (i.e. neighbors) that two nodes have in common,
aiming at capturing the notion that two disconnected elements who share neighbors would be "introduced"
to each other. This feature has been computed in the context of collaboration networks, allowing to verify a positive correlation between the number of shared neighbors of two given nodes $v$ and $v'$ at time $n$, and the probability that $v$ and $v'$ will collaborate at
a posterior time~\cite{newman2001clustering}. As an example, Figure \ref{fig:Example-of-Topological}
presents the formula for \emph{Common Neighbors} along with an example
for packages of Apache Derby (version 10.8.3.0). The feature is
computed for packages \texttt{\small org.apache.derby.impl.sql.execute.rts}
and \texttt{\small org.apache.derby.impl.sql.catalog} (marked in bold in the figure). According to the corresponding dependency graph, both packages have the one and same neighbor. As a
result, the \emph{Common Neighbors} score for the analyzed packages
is $1$. Along this line, some researchers have also studied structural
similarity metrics for source code entities~\cite{seke2013b}, such
as: \emph{Kulczynski}, and \emph{Russell-Rao}, among others. 

\subsubsection*{Content-based features} 
\begin{figure*}
\begin{centering}
\includegraphics[width=17cm,height=7.5cm]{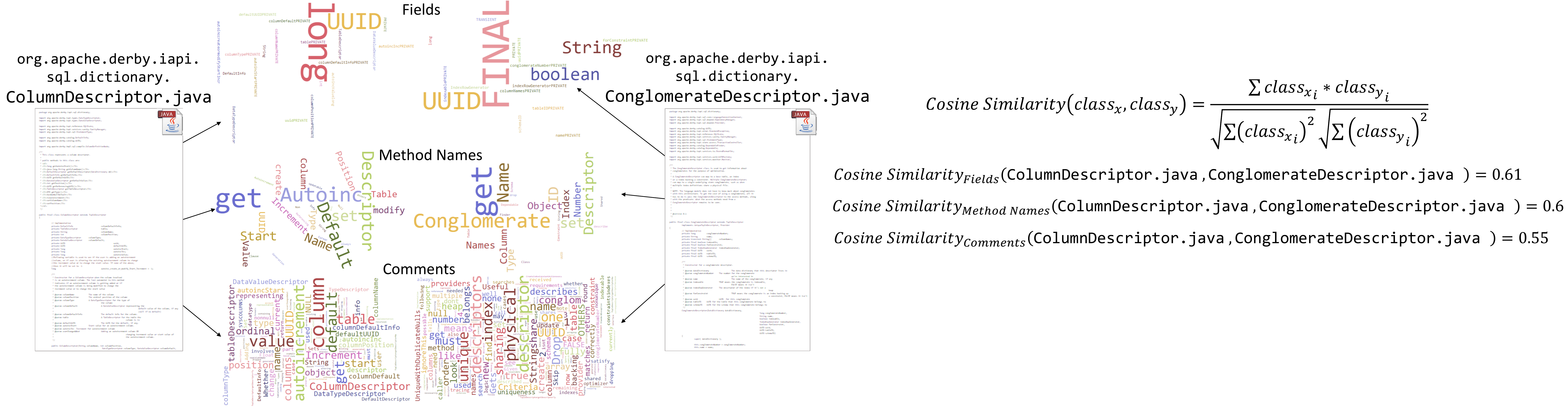}
\par\end{centering}

\caption{\label{fig:Example-of-Content-based}Content-based Representation
of Classes and Computation of Cosine Similarity Scores}
\end{figure*}

These features are an alternative (and complementary) similarity criterion to topological features. Given two texts, the goal of content-based similarity is to determine how close
or similar they are. For example, one of the simplest strategies
is to compute the lexical overlap between texts. Natural language processing routines are used to transform texts into their bag-of-words representations~\cite{Salton:1986:IMI:576628}, which
can be built by considering different aspects of the original texts.
For instance, representations could be restricted to only the appearing
nouns, adjective or verbs, or could choose to remove all punctuation.
In our domain, we can think of each Java class $c$ as a bag-of-words
containing the most representative tokens that characterize its source
code. In this regard, Figure~\ref{fig:Example-of-Content-based}
shows an example of three possible bag-of-words representations
of two Apache Derby classes, either considering the name of the field
attributes of the classes, the name of the declared methods, or the class
comments and documentation. The bag-of-words class representations can be used to assess the similarity among the classes. A common similarity metric for two texts is the \emph{Cosine Similarity}, whose formula is shown in the figure, and computes a score between $0$ and $1$. It is worth noting that each bag-of-words representation could lead to different similarity
scores, as the example shows. Although the presented bag-of-word representation is defined for classes, it can easily be extended to packages. In such a case, the bag-of-words representation of a package $p$ is defined as the (recursive) union of the bag-of-word representations of all the classes (and nested sub-packages) contained by $p$. 

Although intuitive, the homophily principle does not always hold for
complex networks, such as those based on software-related dependencies~\cite{Zhou:2014:BPM:2671850.2671886}.
For instance, two similar packages can intentionally be designed to
not become dependent on each other, based on business logic or modularity
reasons. 
On the contrary, dependencies might still appear between
dissimilar packages. Thus, approaches being able to learn "exceptions"
to homophily are necessary. An interesting approach is to cast the
LP task as a \emph{classification problem} in which a prediction model
is built based on graph information. In this setting, the classification model learns about
both the existence and absence of relations between the different
pairs of nodes in the dependency graph. Furthermore, this approach allows to take the history of the dependency graph into account
(i.e. the graphs corresponding to previous system versions). 

In order to train a classification model, the dependency graph is
converted into a set of instances, called \emph{dataset}, which serves
as the input of the classifier. Each instance consists of a given
pair of nodes, a list of features characterizing the pair, and a label
that indicates whether a dependency exists between the nodes or not.
Those pairs of connected nodes are said to belong to the positive
class, while those pairs of unconnected nodes are said to belong to the negative
class. Table~\ref{tab:Instance-Example} presents the instance-based
representation for some of the pairs of packages depicted in Figure~\ref{fig:Example-of-Topological}.
As it can be observed, instances are represented by two features:
the \emph{Common Neighbors} score and the \emph{Cosine Similarity} of the comment
bag-of-words representation of packages. 
\begin{table}
\begin{centering}
{\scriptsize \setlength{\tabcolsep}{2pt}}%
\begin{tabular}{|>{\centering}p{0.22\columnwidth}|>{\centering}p{0.10\columnwidth}|>{\centering}p{0.10\columnwidth}|>{\centering}p{0.10\columnwidth}|}
\hline 
\emph{\scriptsize Source}\\
\emph{\scriptsize Target} & \emph{\scriptsize Source depends on target} & \emph{\scriptsize Common Neighbors} & \emph{\scriptsize Cosine Similarity Comments}\tabularnewline
\hline 
\hline 
\emph{\scriptsize org.apache.derby.impl.sql}\\
\emph{\scriptsize org.apache.derby.catalog} & {\scriptsize true} & {\scriptsize 0.353} & {\scriptsize 0.835}\tabularnewline
\hline 
\emph{\scriptsize org.apache.derby.impl.sql}\\
\emph{\scriptsize org.apache.derby.impl.sql.catalog} & {\scriptsize false} & {\scriptsize 0.618} & {\scriptsize 0.877}\tabularnewline
\hline 
\emph{\scriptsize org.apache.derby.catalog}\\
\emph{\scriptsize org.apache.derby.impl.sql.execute} & {\scriptsize true} & {\scriptsize 0.389} & {\scriptsize 0.870}\tabularnewline
\hline 
\emph{\scriptsize org.apache.derby.impl.sql.catalog}\\
\emph{\scriptsize org.apache.derby.catalog} & {\scriptsize true} & {\scriptsize 0.385} & {\scriptsize 0.877}\tabularnewline
\hline 
\emph{\scriptsize ...} & {\scriptsize ...} & {\scriptsize ...} & {\scriptsize ...}\tabularnewline
\hline 
\emph{\scriptsize org.apache.derby.impl.sql.execute}\\
\emph{\scriptsize org.apache.derby.impl.sql.catalog} & {\scriptsize false} & {\scriptsize 0.605} & {\scriptsize 0.939}\tabularnewline
\hline 
\emph{\scriptsize org.apache.derby.impl.sql.execute}\\
\emph{\scriptsize org.apache.derby.impl.sql.execute.rts} & {\scriptsize true} & {\scriptsize 0.171} & {\scriptsize 0.487}\tabularnewline
\hline 
\end{tabular}
\par\end{centering}

\caption{\label{tab:Instance-Example}Example of Dataset (Apache Derby, version 10.8.3.0)}

\end{table}

In~\cite{diazpace_icsa_2018}, we assessed the predictive power of LP techniques for inferring (individual) package dependencies using topological information from system versions. The results showed that
classification models can provide reasonable predictions, assuming certain conditions in the pairs of versions
used for the predictions (e.g., the consecutive versions have almost
the same number of packages, a certain percentage of dependencies is added
in the next version). Nonetheless, opportunities for improving the classification were identified. In particular, we hypothesize 
that content-based features can boost the performance of the trained
classification models. To this end, instances in this work are
characterized by means of both topological and content-based features.
As regards topological features, we considered: \emph{Adamic-Adar}, \emph{Common
Neighbours}, \emph{Resource Allocation} and \emph{Sørensen},
as well as the best performing metrics in~\citep{seke2013b}, namely: \emph{Kulczynski},
\emph{RelativeMatching} and \emph{RusselRao}. For
content-based features, we included the \emph{Cosine Similarity} scores for the
different bag-of-words representations previously defined, as well
as representations considering method usage and variable definitions. 
In practice, a dataset of instances can initially include many different features (both topological and content-based ones), and then feature selection techniques  
~\cite{Forman:2003:EES:944919.944974} can be applied for (automatically) determining the subset of features being most relevant (or providing more information) for the classification task. 

\section{Approach}
\label{sec:approach} 
\begin{figure*}
\begin{centering}
\includegraphics[height=0.35\textheight]{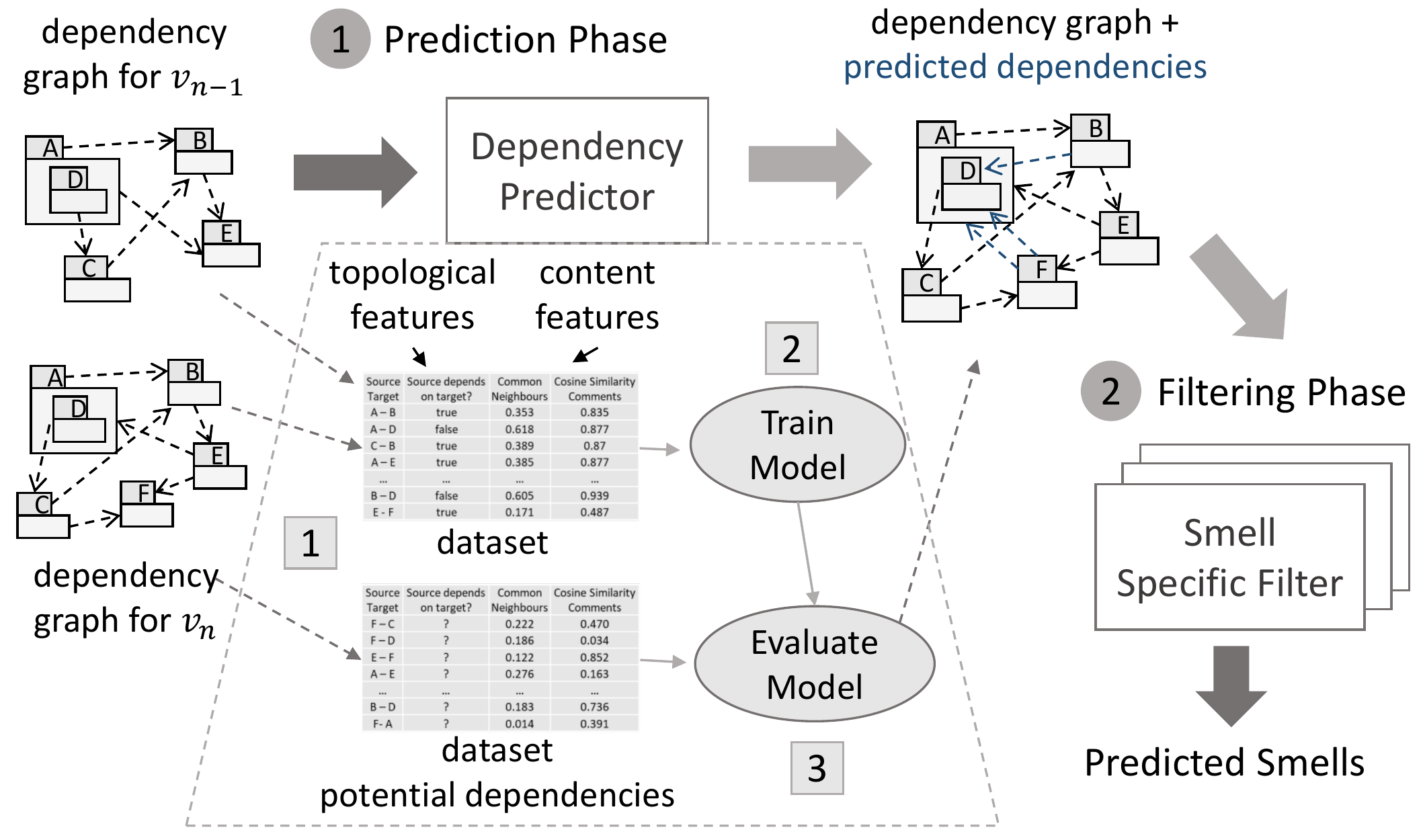}
\par\end{centering}

\caption{\label{fig:Overview-of-the}Two-phase Approach for Predicting Architectural Smells}

\end{figure*}

Our approach works in two phases: first, it seeks to predict the appearance of new dependencies in the next system version, and then it filters them according to the characteristics of specific types of smells. Figure~\ref{fig:Overview-of-the} depicts an overview of the main building blocks of the approach. Although the approach supports different architectural smells, in this work we customize the second phase for the CD and HLD smells.

Initially, there is a \textit{prediction phase} in which individual dependencies are inferred based on training a binary classification model. To do so, the dependency graphs corresponding to the current and previous versions (noted as $v_{n}$ and $v_{n-1}$, respectively) are required as inputs. The output of this phase is the set of dependencies that are likely to appear in the next system version $v_{n+1}$. This phase is \textit{smell-independent}, in the sense that only identifies dependencies that might prefigure smells in the second phase.
The prediction phase internally involves 3 steps. In step $1$, an instance-based representation of system versions (as presented in Section~\ref{sec:link-prediction}) is constructed, based on both topological and content-based features. The goal of the classifier is to predict which new dependencies are likely to appear; hence, those pairs of nodes already connected in the graph are considered as the \textit{positive class}, while those pairs without dependencies are considered as the \textit{negative class}. In step $2$,
the classification model is built. As the figure shows, the training
set comprises instances belonging to two system versions: i) existing
dependencies in $v_{n-1}$ (as instances of the positive class), ii) missing dependencies in $v_{n-1}$ (as instances of the negative class), and iii) existing dependencies in $v_{n}$. It is worth noting that the information from the system versions serves to train the classifier for properly learning instances of both the positive and negative
classes. This mechanism allows us to include information of dependencies in $v_{n-1}$ that are guaranteed not to appear
in the next version ($v_{n}$). If we would only consider one system, no information regarding the negative class
could be included in the model training, as it would not possible to guarantee that those dependencies are not going to appear in the next version. 

Once the model is trained, step $3$ predicts which dependencies could
appear in $v_{n+1}$. Note also that the prediction might face the
case of dependencies in $v_{n+1}$ appearing between packages that
did not originally exist in $v_{n}$ (e.g., due to the creation of
new packages). Although it is possible to predict that an existing
package will depend on an unobserved package, we cannot
determine what that unobserved package will be. In this regard, the
trained classifier predicts whether the pairs of nodes that are not
dependent on each other (i.e. they are unconnected) in $v_{n}$
might become dependent in $v_{n+1}$. This means that only potential dependencies
considering the packages already existing in $v_{n}$ are considered. 

The fact that the classifier predicts if an individual dependency is likely to appear is not enough to actually predict the appearance of an architectural smell, since not every predicted dependency might cause a smell to emerge. Usually, an emerging smell is the result of a group of predicted and existing dependencies. To this end, in the second phase, predicted dependencies undergo a filtering process according to the type of smell at hand. As shown in the figure, this \textit{filtering phase} requires the creation of filters for each smell type. The current filters for CD and HLD are described below.  

\subsubsection*{Cycle Filter} It considers only predicted
dependencies that lead to the closure of new cycles in $v_{n+1}$ (those cycles must not exist in previous versions). The predicted dependencies are considered all altogether, and simultaneously added to graph at $v_{n+1}$, before checking for cycles.

\subsubsection*{Hub Filter}\label{sec:hub-filter} 
\begin{figure}
\begin{centering}
\includegraphics[width=0.95\columnwidth]{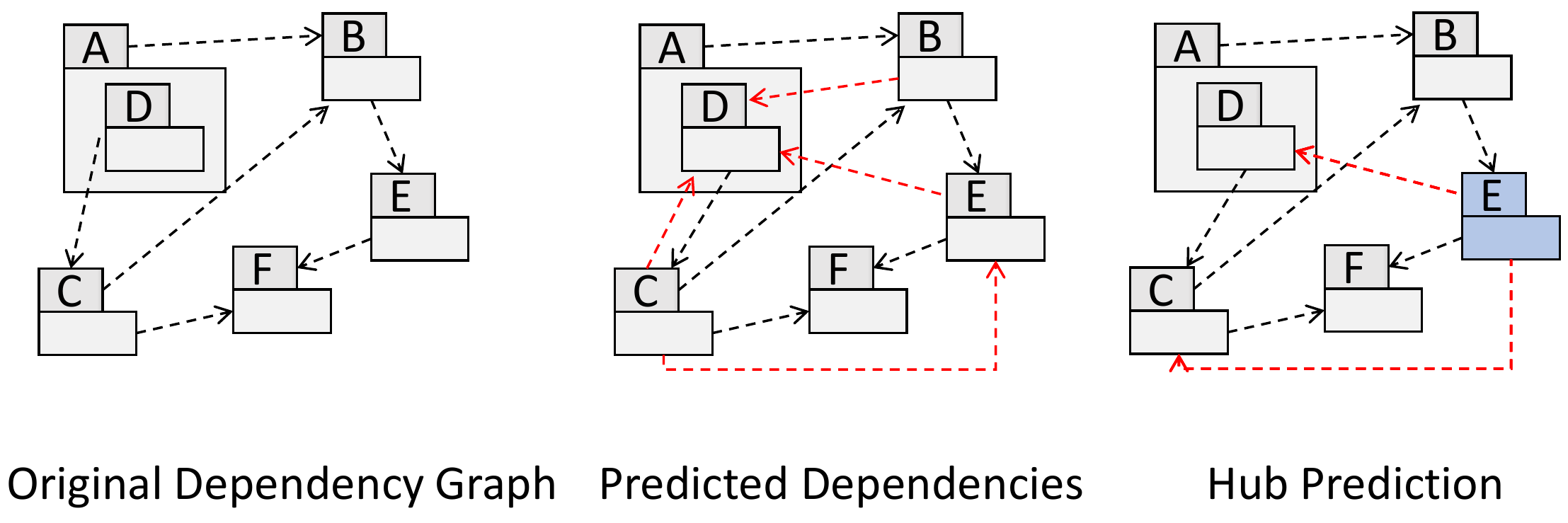}
\par\end{centering}

\caption{\label{fig:Hub-Variants-Example}Hub Variants Example}

\end{figure}

It considers only the nodes incidental to the predicted dependencies that fit with the hub definition of Section~\ref{sec:hld}. This process is exemplified in Figure~\ref{fig:Hub-Variants-Example}.
The default strategy\footnote{Other variants are possible, which are not covered due to space reasons} works as following: for each node that is incidental to at least one predicted dependency, all its actual and predicted dependencies are analyzed together to compute the hub score of the node. For example, for computing the hub score of package E, both dependencies between E and D, and C and E are jointly considered in the analysis. Note that this strategy allows the detection of those nodes becoming hubs due to the addition of new dependencies, but disregards nodes that might become hubs due to changes in the overall structure of the dependency graph (i.e. nodes for which no new dependencies are added).

\section{Study Settings}

In order to assess the performance of the proposed approach for predicting architectural smells, we exercised it with two Java open-source systems: Apache Derby\footnote{\url{https://db.apache.org/derby/}}
($\sim$40.30 KLOC) and Apache Ant\footnote{\url{https://ant.apache.org/}}
($\sim$59.79 KLOC), which have been also used by others in the literature \cite{DBLP:conf/icsm/FontanaPRZ16}. We selected systems with more than 10 major releases, and more than 10 contributors each. Derby versions correspond to the period 2008-2014, while Ant versions correspond to 2003-2018.

Table~\ref{tab:system-versions-1} summarizes the main characteristics
of the two systems. Note that the number of changes with respect
to the previous version might not match the absolute difference between
the two versions. This is due to the fact that the noted differences
correspond to changes/smells that could be predicted. Instead, changes caused by the addition of new packages are disregarded. The dependency graphs
for the versions were extracted with the
CDA tool\footnote{\url{http://www.dependency-analyzer.org}} that performs a static analysis of binary code.  For the purpose of this work, class dependencies were ignored. To
obtain meaningful predictions with our approach, the pairs of consecutive versions to
analyze ($v_{n-1}$, $v_{n}$) need to present changes regarding the existence of cycles or hubs, which implies the addition of packages dependencies from one version to the next one. As a result of this criterion, the subset of versions for which we made predictions are highlighted in bold (for new cycles) and italic (for new hubs) in the table. 
Cycles were detected by means of the Arcan\footnote{http://essere.disco.unimib.it/wiki/arcan} tool.
Finally, content information of classes and packages was extracted
by parsing the source code with Java Parser\footnote{http://javaparser.org/} and then computing the metrics with the Java String Similarity\footnote{https://github.com/tdebatty/java-string-similarity} library. All the content-based features presented in Section \ref{sec:link-prediction} were computed, and then the information gain method  \cite{Forman:2003:EES:944919.944974} for feature selection was applied. 


\begin{table}
\begin{centering}
{\scriptsize }%
\begin{tabular}{|>{\centering}p{0.15\columnwidth}|c|c|>{\centering}p{0.1\columnwidth}|>{\centering}p{0.065\columnwidth}|>{\centering}p{0.06\columnwidth}|>{\centering}p{0.10\columnwidth}|>{\centering}p{0.07\columnwidth}|}
\hline
\hline
 & {\scriptsize \#c} & {\scriptsize \#p} & {\scriptsize \#deps} & {\scriptsize \#cycles} & {\scriptsize cycle length} & {\scriptsize \#hubs} & {\scriptsize hub degree}\tabularnewline
\hline 
\hline 
{\scriptsize \centering\linespread{0.8}derby 10.3.3.0} & {\scriptsize 1261} & {\scriptsize 81} & {\scriptsize 698} & {\scriptsize 192} & {\scriptsize 10.80} & {\scriptsize 22} & {\scriptsize 28.68}\tabularnewline
\hline
{\scriptsize \centering\linespread{0.8}derby 10.4.1.0} & {\scriptsize 1321} & {\scriptsize 94} & {\scriptsize 757, +61, -2} & {\scriptsize 196, +4} & {\scriptsize 10.55} & {\scriptsize 26 - +4} & {\scriptsize 28.34}\tabularnewline
\hline
{\scriptsize \centering\linespread{0.8}derby 10.4.2.0 } & {\scriptsize 1322} & {\scriptsize 94} & {\scriptsize 757} & {\scriptsize 196} & {\scriptsize 10.55} & {\scriptsize 26} & {\scriptsize 28.34}\tabularnewline
\hline 
{\scriptsize \centering\linespread{0.8}\textbf{\emph{derby 10.5.1.1}}} & \textbf{\emph{\scriptsize 1344}} & \textbf{\emph{\scriptsize 96}} & \textbf{\emph{\scriptsize 767, +10}} & \textbf{\emph{\scriptsize 234, +4}} & {\scriptsize 11.59} & \textbf{\emph{\scriptsize 28, +2}} & {\scriptsize 27.67}\tabularnewline
\hline
{\scriptsize \centering\linespread{0.8}\textbf{derby 10.5.3.0}} & \textbf{\scriptsize 1344} & \textbf{\scriptsize 96} & \textbf{\scriptsize 768}\textbf{\scriptsize +1} & \textbf{\scriptsize 234} & {\scriptsize 11.59} & \textbf{\scriptsize 28} & {\scriptsize 27.71}\tabularnewline
\hline
{\scriptsize \centering\linespread{0.8}\textbf{\emph{derby 10.6.1.0}}} & \textbf{\emph{\scriptsize 1387}} & \textbf{\emph{\scriptsize 98}} & \textbf{\emph{\scriptsize 804 , +36}} & \textbf{\emph{\scriptsize 254, +6}} & {\scriptsize 12.99} & \textbf{\emph{\scriptsize 29, +3}} & {\scriptsize 28.34}\tabularnewline
\hline
{\scriptsize \centering\linespread{0.8}\textbf{derby 10.6.2.1}} & \textbf{\scriptsize 1387} & \textbf{\scriptsize 98} & \textbf{\scriptsize 805}\textbf{\scriptsize +1} & \textbf{\scriptsize 255}\textbf{\scriptsize +1} & {\scriptsize 13.02} & \textbf{\scriptsize 29} & {\scriptsize 28.34}\tabularnewline
\hline
{\scriptsize \centering\linespread{0.8}\textbf{derby 10.7.1.1}} & \textbf{\scriptsize 1389} & \textbf{\scriptsize 98} & \textbf{\scriptsize 807}\textbf{\scriptsize +4, -2} & \textbf{\scriptsize 257}\textbf{\scriptsize +2} & {\scriptsize 12.98} & \textbf{\scriptsize 29} & {\scriptsize 28.44}\tabularnewline
\hline
{\scriptsize \centering\linespread{0.8}\textbf{\emph{derby 10.8.1.2}}} & \textbf{\emph{\scriptsize 1395}} & \textbf{\emph{\scriptsize 97}} & \textbf{\emph{\scriptsize 837, +31, -1}} & \textbf{\emph{\scriptsize 305, +22}} & {\scriptsize 15.17} & \textbf{\emph{\scriptsize 30, +1}} & {\scriptsize 30.03}\tabularnewline
\hline
{\scriptsize \centering\linespread{0.8}derby 10.8.2.2} & {\scriptsize 1395} & {\scriptsize 96} & {\scriptsize 838, +2, -1} & {\scriptsize 305} & {\scriptsize 15.17} & {\scriptsize 30} & {\scriptsize 30.03}\tabularnewline
\hline
{\scriptsize \centering\linespread{0.8}\textbf{derby 10.8.3.0}} & \textbf{\scriptsize 1395} & \textbf{\scriptsize 96} & \textbf{\scriptsize 841}\textbf{\scriptsize +3} & \textbf{\scriptsize 306}\textbf{\scriptsize +1} & {\scriptsize 15.13} & \textbf{\scriptsize 30} & {\scriptsize 30.06}\tabularnewline
\hline
{\scriptsize \centering\linespread{0.8}\textbf{\emph{derby 10.9.1.0}}} & \textbf{\emph{\scriptsize 1406}} & \textbf{\emph{\scriptsize 96}} & \textbf{\emph{\scriptsize 851, +20, -10}} & \textbf{\emph{\scriptsize 280, +5}} & {\scriptsize 13.43} & \textbf{\emph{\scriptsize 29, +1}} & {\scriptsize 30.62}\tabularnewline
\hline
{\scriptsize \centering\linespread{0.8}\textbf{\emph{derby 10.10.1.1}}} & \textbf{\emph{\scriptsize 1453}} & \textbf{\emph{\scriptsize 100}} & \textbf{\emph{\scriptsize 938, +89, -2}} & \textbf{\emph{\scriptsize 291, +10}} & {\scriptsize 13.32} & \textbf{\emph{\scriptsize 29, -1}} & {\scriptsize 32.89}\tabularnewline
\hline
{\scriptsize \centering\linespread{0.8}derby 10.10.2.0} & {\scriptsize 1454} & {\scriptsize 100} & {\scriptsize 938} & {\scriptsize 291} & {\scriptsize 13.32} & {\scriptsize 29} & {\scriptsize 32.89}\tabularnewline
\hline
\multicolumn{1}{>{\centering}p{0.09\columnwidth}}{} & \multicolumn{1}{c}{} & \multicolumn{1}{c}{} & \multicolumn{1}{>{\centering}p{0.1\columnwidth}}{} & \multicolumn{1}{>{\centering}p{0.065\columnwidth}}{} & \multicolumn{1}{>{\centering}p{0.06\columnwidth}}{} & \multicolumn{1}{>{\centering}p{0.05\columnwidth}}{} & \multicolumn{1}{>{\centering}p{0.07\columnwidth}}{}\tabularnewline
\hline
{\scriptsize \centering\linespread{0.8}ant 1.5.2} & {\scriptsize 297} & {\scriptsize 21} & {\scriptsize 71} & {\scriptsize 21} & {\scriptsize 3.57} & {\scriptsize 7} & {\scriptsize 13.71}\tabularnewline
\hline
{\scriptsize \centering\linespread{0.8}ant 1.5.3-1} & {\scriptsize 297} & {\scriptsize 21} & {\scriptsize 71} & {\scriptsize 21} & {\scriptsize 3.57} & {\scriptsize 7} & {\scriptsize 13.71}\tabularnewline
\hline
{\scriptsize \centering\linespread{0.8}\textbf{\emph{ant 1.6.0}}} & \textbf{\emph{\scriptsize 352}} & \textbf{\emph{\scriptsize 24}} & \textbf{\emph{\scriptsize 90, +20, -1}} & \textbf{\emph{\scriptsize 30, +1}} & {\scriptsize 3.73} & \textbf{\emph{\scriptsize 9, +2}} & {\scriptsize 14.22}\tabularnewline
\hline
{\scriptsize \centering\linespread{0.8}ant 1.6.1} & {\scriptsize 353} & {\scriptsize 24} & {\scriptsize 90} & {\scriptsize 30} & {\scriptsize 3.73} & {\scriptsize 9} & {\scriptsize 14.22}\tabularnewline
\hline 
{\scriptsize \centering\linespread{0.8}\textbf{ant 1.6.2}} & \textbf{\scriptsize 369} & \textbf{\scriptsize 24} & \textbf{\scriptsize 92}\textbf{\scriptsize +2} & \textbf{\scriptsize 43}\textbf{\scriptsize +2} & {\scriptsize 4.12} & \textbf{\scriptsize 9} & {\scriptsize 14.67}\tabularnewline
\hline
{\scriptsize \centering\linespread{0.8}\textbf{ant 1.6.3}} & \textbf{\scriptsize 380} & \textbf{\scriptsize 25} & \textbf{\scriptsize 97}\textbf{\scriptsize +5} & \textbf{\scriptsize 43}\textbf{\scriptsize +1} & {\scriptsize 4.70} & \textbf{\scriptsize 9} & {\scriptsize 15.33}\tabularnewline
\hline
{\scriptsize \centering\linespread{0.8}ant 1.6.4} & {\scriptsize 380} & {\scriptsize 25} & {\scriptsize 97} & {\scriptsize 43} & {\scriptsize 4.70} & {\scriptsize 9} & {\scriptsize 15.33}\tabularnewline
\hline
{\scriptsize \centering\linespread{0.8}ant 1.6.5} & {\scriptsize 380} & {\scriptsize 25} & {\scriptsize 97} & {\scriptsize 43} & {\scriptsize 4.70} & {\scriptsize 9} & {\scriptsize 15.33}\tabularnewline
\hline
{\scriptsize \centering\linespread{0.8}\textbf{\emph{ant 1.7.1}}} & \textbf{\emph{\scriptsize 502}} & \textbf{\emph{\scriptsize 29}} & \textbf{\emph{\scriptsize 137, +46, -6}} & \textbf{\emph{\scriptsize 63, +5}} & {\scriptsize 5.10} & \textbf{\emph{\scriptsize 12, +1}} & {\scriptsize 17.42}\tabularnewline
\hline
{\scriptsize \centering\linespread{0.8}ant 1.8.0} & {\scriptsize 557} & {\scriptsize 30} & {\scriptsize 148, +12, -1} & {\scriptsize 69, +4} & {\scriptsize 5.91} & {\scriptsize 12} & {\scriptsize 18.58}\tabularnewline
\hline
{\scriptsize \centering\linespread{0.8}ant 1.9.3} & {\scriptsize 772} & {\scriptsize 61} & {\scriptsize 282, +134} & {\scriptsize 101, +3} & {\scriptsize 6.25} & {\scriptsize 12, +5} & {\scriptsize 17.38}\tabularnewline
\hline
{\scriptsize \centering\linespread{0.8}ant 1.9.4} & {\scriptsize 774} & {\scriptsize 61} & {\scriptsize 283, +1} & {\scriptsize 101} & {\scriptsize 6.25} & {\scriptsize 24} & {\scriptsize 17.42}\tabularnewline
\hline
{\scriptsize \centering\linespread{0.8}ant 1.9.5} & {\scriptsize 776} & {\scriptsize 61} & {\scriptsize 283} & {\scriptsize 101} & {\scriptsize 6.25} & {\scriptsize 24} & {\scriptsize 17.42}\tabularnewline
\hline 
{\scriptsize \centering\linespread{0.8}ant 1.9.9} & {\scriptsize 780} & {\scriptsize 61} & {\scriptsize 282, -1} & {\scriptsize 100, -1} & {\scriptsize 6.29} & {\scriptsize 24} & {\scriptsize 17.38}\tabularnewline
\hline
{\scriptsize \centering\linespread{0.8}ant 1.10.0} & {\scriptsize 782} & {\scriptsize 61} & {\scriptsize 284, +2} & {\scriptsize 100} & {\scriptsize 6.29} & {\scriptsize 24} & {\scriptsize 17.46}\tabularnewline
\hline
{\scriptsize \centering\linespread{0.8}ant 1.10.1} & {\scriptsize 782} & {\scriptsize 61} & {\scriptsize 284} & {\scriptsize 100} & {\scriptsize 6.29} &{\scriptsize 24} & {\scriptsize 17.46}\tabularnewline
\hline
{\scriptsize \centering\linespread{0.8}ant 1.10.2} & {\scriptsize 784} & {\scriptsize 61} & {\scriptsize 283, +1, -2} & {\scriptsize 100} & {\scriptsize 6.29} & {\scriptsize 24} & {\scriptsize 17.42}\tabularnewline
\hline
{\scriptsize \centering\linespread{0.8}ant 1.10.3} & {\scriptsize 784} & {\scriptsize 61} & {\scriptsize 283} & {\scriptsize 100} & {\scriptsize 6.29} & {\scriptsize 24} & {\scriptsize 17.42}\tabularnewline

\hline 
\end{tabular}
\par\end{centering}{\scriptsize \par}

\begin{centering}
\begin{tabular}{>{\centering}p{0.95\columnwidth}}
{\scriptsize where }\emph{\scriptsize \#c}{\scriptsize{} indicates
number of classes, }\emph{\scriptsize \#p}{\scriptsize{} number of
packages, }\emph{\scriptsize \#deps}{\scriptsize{} number of dependencies,
and }\emph{\scriptsize +}{\scriptsize{} and }\emph{\scriptsize -}{\scriptsize{}
indicate the number of changes regarding the previous version}\tabularnewline
\end{tabular}
\par\end{centering}

\caption{\label{tab:system-versions-1}General Characteristics of the Selected
Systems}
\end{table}

As previously mentioned, those dependencies that do not yet exist
in $v_{n}$ are used as the input of the approach and thus constitute the test set. To assess the performance, $v_{n+1}$ is used
to determine whether a dependency should be predicted. As a result, the test set follows the real class distribution of
the system versions, i.e. negative instances are not under-sampled.
Classification was performed using the Weka\footnote{ https://www.cs.waikato.ac.nz/ml/weka}
implementation of the SVM algorithm (Support Vector Machine), parametrized with a RBF kernel,
which is useful for datasets with few instances of one of the classes.
Performance was assessed by considering the traditional precision (i.e., the ratio between the number of actual discovered dependencies and the total number of predictions) and
recall (i.e., the ratio between the actual discovered dependencies and the total number of actual dependencies) metrics. In principle, both recall and precision are relevant performance metrics for the predictions; however, we are more interested in recall than in precision, because this would indicate that all the smells are effectively identified (for the next version). Eventually, some smells can be mistakenly predicted and negatively impact on precision. If those smells are a small fraction, they could be discarded with a manual analysis of the outputs (of the approach) by developers.

\section{Evaluation} 
\label{sec:evaluation-results} 

In this section, we discuss the predictions of the first phase of the approach, and how they influenced the results of the second phase. The cases of CD and HLD are presented separately. In the reported results, for each pair ($X$ axis), the versions
represent the span for the predictions, e.g., $v_{1}$-{$v_{2}$}-{$v_{3}$}
means that $v_{1}$ and $v_{2}$ served to train the model for predicting
new architectural smells in $v_{3}$. We refer to a quasi-cycle (or a quasi-hub) as a configuration of dependencies in the current system version that does not form a cycle (or a hub) yet, but upon the inclusion of predicted dependencies, it might become a cycle (or a hub) in the next version.

\subsubsection*{Dependency Prediction Phase}
\begin{figure*}
\begin{centering}
\subfloat[Apache Derby]{\includegraphics[width=0.95\textwidth]{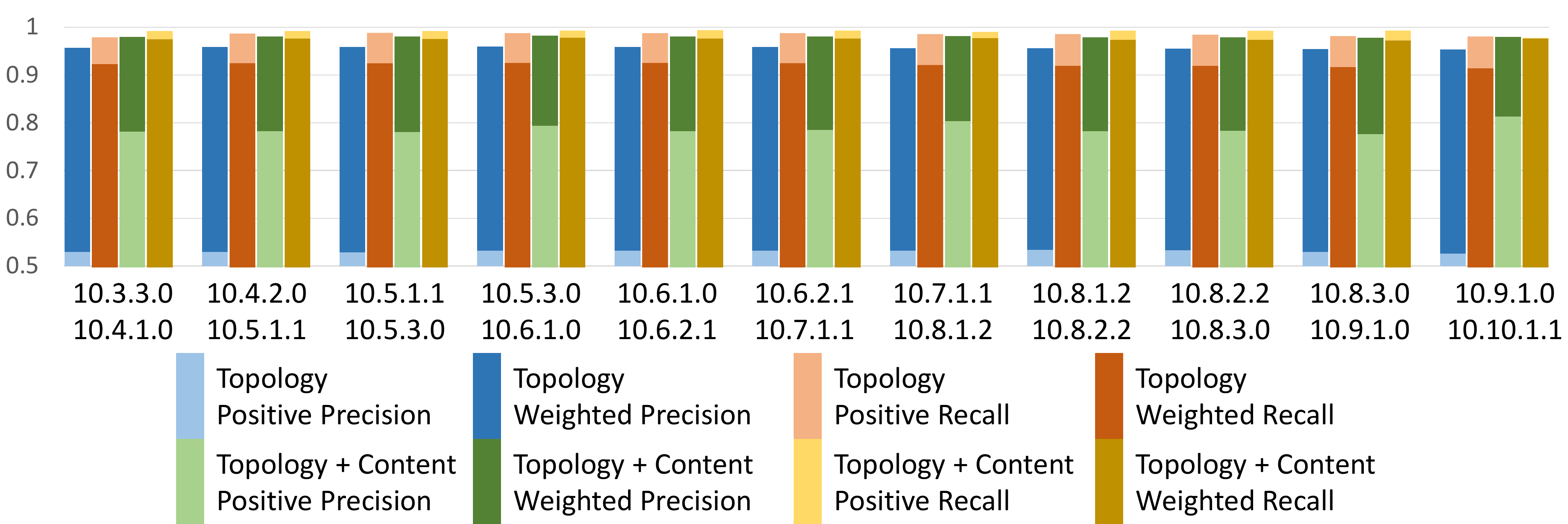}

}\\
\subfloat[Apache Ant]{\includegraphics[width=0.95\textwidth]{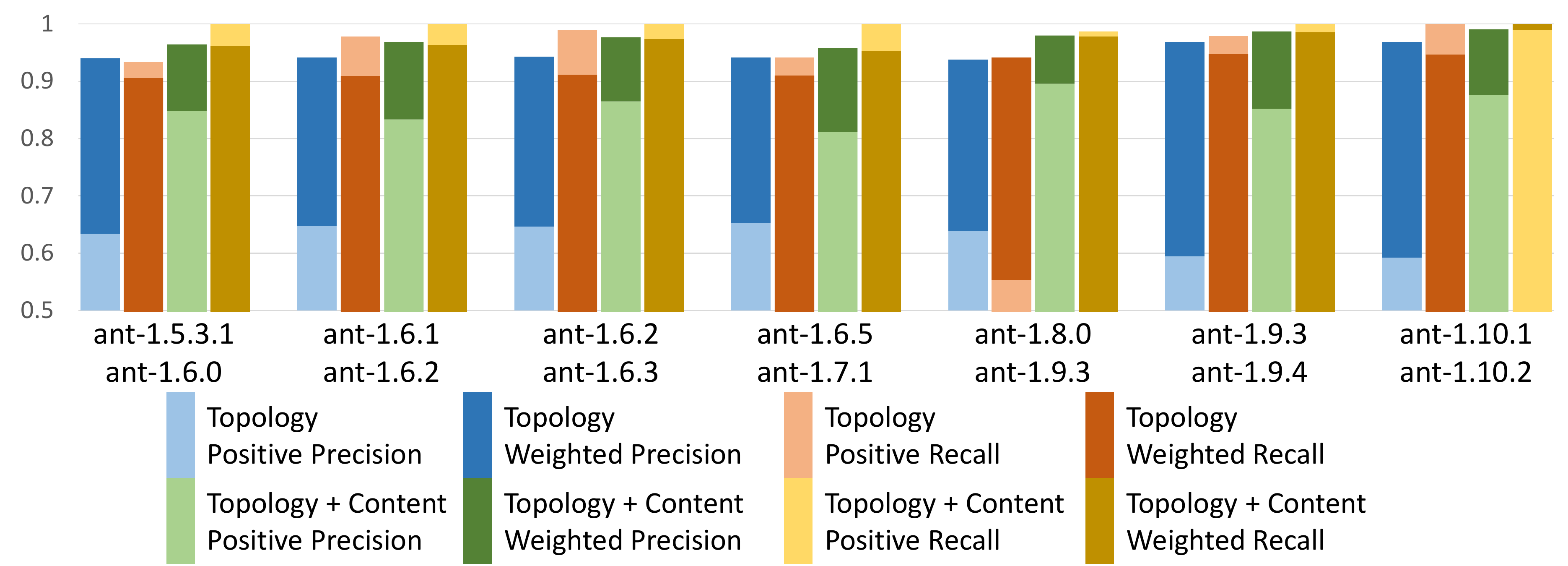}}\caption{\label{fig:classification-results}Results for the Dependency Prediction
Phase}

\par\end{centering}

\end{figure*}

Figure~\ref{fig:classification-results} presents the results 
of the first phase for the two systems,
when considering: i) topological features only, and ii) a combination of topological and content-based features. We computed precision and recall for the positive
class (i.e., existing dependencies) and also a weighted F-Measure
considering both classes (i.e., existing and non-existing dependencies). In practice, the positive class is the most relevant indicator, but we believe the weighted indicator provides a context for the former.
Results are presented for those sets of versions in which new dependencies between already existing packages were added. As it can be observed, adding content-based features to the classification model increased the quality of the predicted dependencies. In particular, the improvement was most noticeable in precision for the positive class. The usage of (only) topological features achieved an average precision value of 0.5 and 0.6 in average for the positive class, while the combination of content-based features reached values of 0.75 and  0.85 in average for the same class for Apache Derby and Apache Ant, respectively. The average improvements were $48\%$ and $32\%$ when analyzing Apache Derby and Apache Ant, respectively. This boosting in the classification also outperforms the initial results (without content-based features) reported in \cite{diazpace_icsa_2018}.
It is worth noting that the highest improvements for Apache Ant were
observed for the pair of versions that yielded the lowest topology results (\emph{ant-1.10.1} - \emph{ant-1.10.2}). This fact highlights the importance of considering additional features for characterizing the dependencies between packages, and hence, having better predictions of their evolution over time.

When considering only topological features, both systems reported good results in cases of a high recall and a moderate precision, which means that the trained model is capable of finding most future dependencies, but it also predicts false dependencies, i.e. dependencies that will not appear in the next version. Interestingly, for some of the Apache Ant versions, recall was perfect. When adding content-based features, we noticed that precision improved (i.e. the false dependencies decreased) while recall improved or remained the same.

In summary, including content-based features in the classification model helped to reduce the number of: i) missed dependencies that should have been predicted, and ii) dependencies that should not have been predicted. Having few mistaken predictions has consequences on the next phase of the approach, as it contributes to the reduction of mistakes in the (final) smell prediction. The average computation time for building the classifiers was around 10-15 seconds for Apache Derby and 1-2 seconds for Apache Ant (both cycles and hubs), on a PC i7-4500U 1.8 GHz. with 8GB RAM - Windows 10 and Java 1.8. The response time of the filters was negligible in the whole pipeline. In general, the computation time is affected by factors such as: number of features, size of the dependency graphs, and distribution of smells in the system versions.

\subsubsection*{Cycle Prediction}\label{sec:cycle-prediction}
\begin{figure}
\begin{centering}
\subfloat[Apache Derby]{\includegraphics[width=0.95\columnwidth]{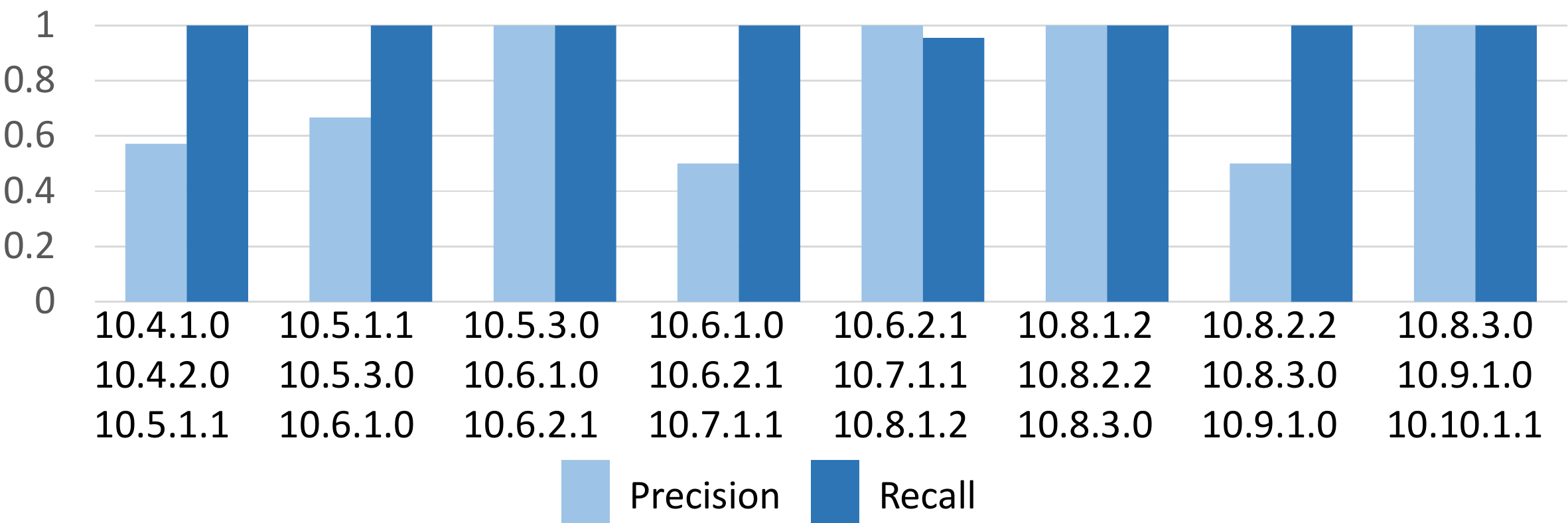}

}\\
\subfloat[Apache Ant]{\includegraphics[width=0.95\columnwidth]{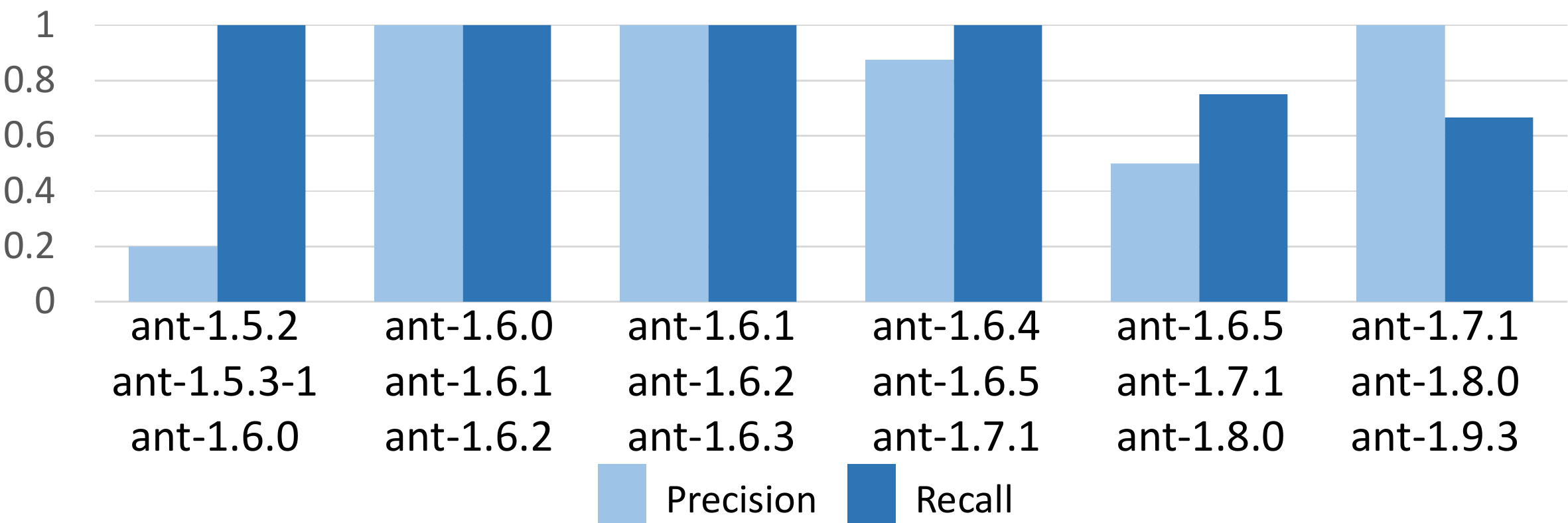}}\caption{\label{fig:Cycle-Prediction-Results}Cycle Prediction Results}

\par\end{centering}

\end{figure}

The values of precision and recall for predicting CD smells are presented
in Figure~\ref{fig:Cycle-Prediction-Results}. As it can be observed,
recall is almost perfect, meaning that almost every
new dependency leading to the closure of a quasi-cycle was found.
On the other hand, precision results indicate that, in addition to
predicting the correct dependencies, some other mistaken dependencies
were also predicted. Nonetheless, even when precision might seem low
(reaching minimum values of $0.5$ and $0.2$ for Apache Derby and
Apache Ant, respectively), at most $5$ mistaken predictions were made.
This amount represents at most a $0.06\%$ or $1.14\%$ of the total
number of dependencies in the corresponding software versions. Therefore,
the incidence of that fraction of mistaken dependencies is low in relation
to the size of the system graphs. The mistaken dependencies
correspond to cases in which a dependency predicted by the first
phase of the approach would potentially close a quasi-cycle, but such a dependency did not actually appear on the next system version. Consequently,
solving this problem would require mechanisms to diminish the influence of mistaken dependencies from the first phase.

\subsubsection*{Hub Prediction}
\begin{figure}
\begin{centering}
\subfloat[Apache Derby]{\includegraphics[width=0.95\columnwidth]{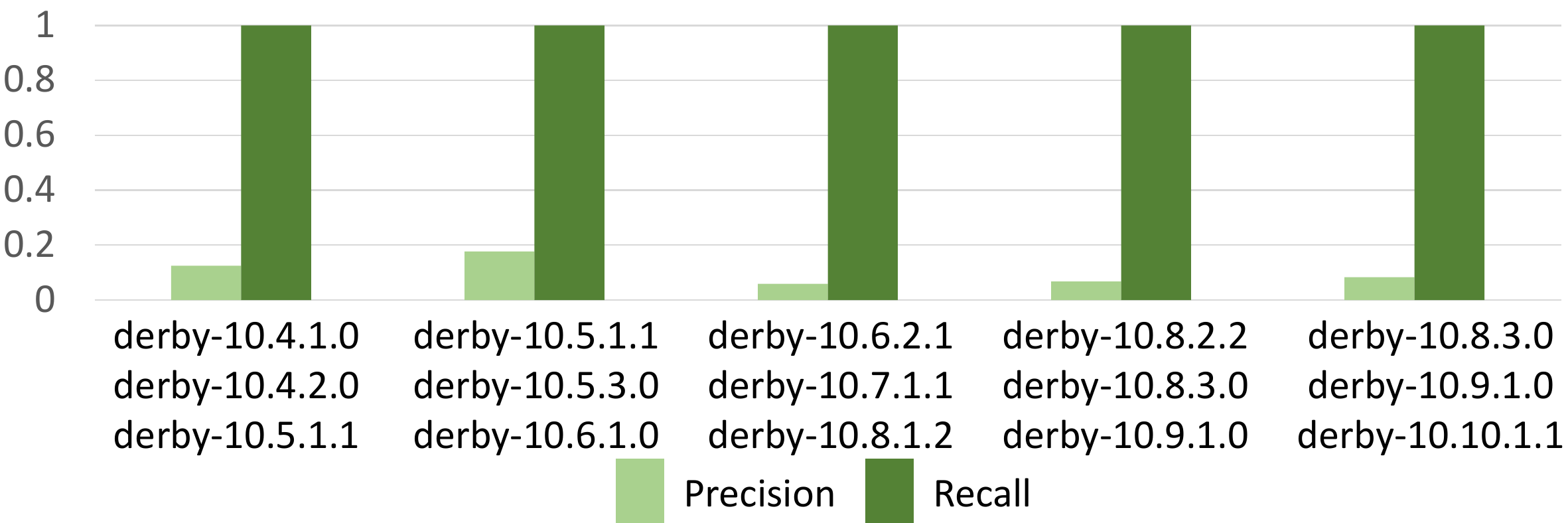}}\\
\subfloat[Apache Ant]{\includegraphics[width=0.95\columnwidth]{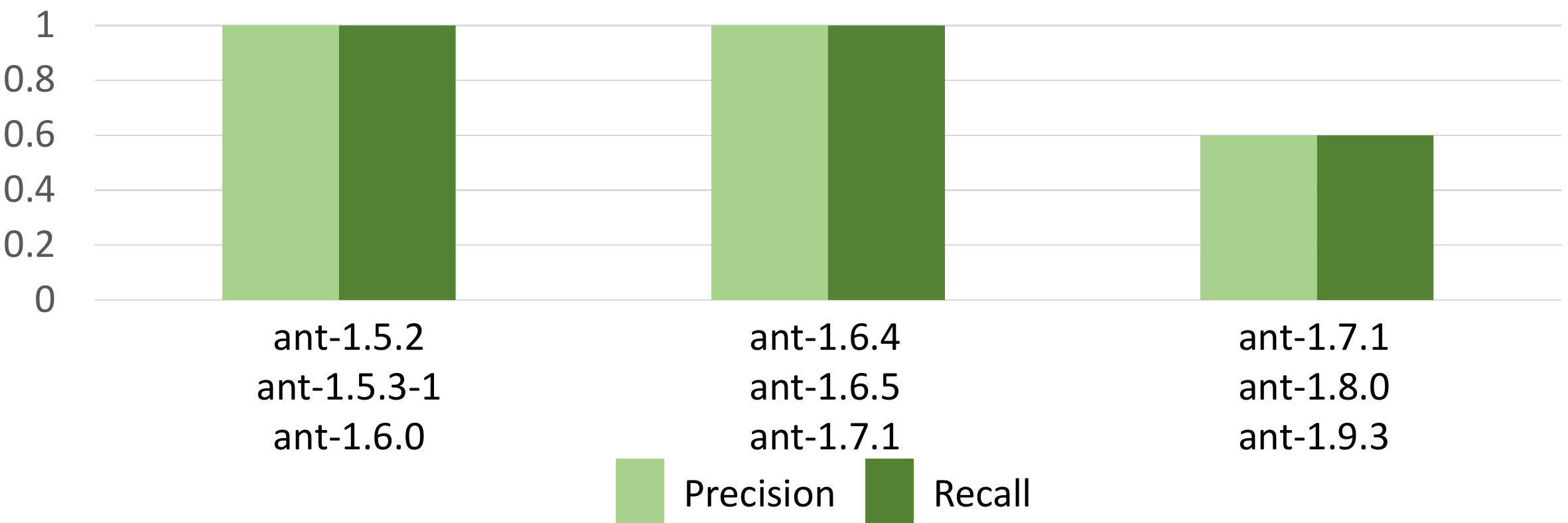}}
\par\end{centering}

\caption{\label{fig:Hub-Prediction-Results}Hub Prediction Results}
\end{figure}

The values of precision and recall for predicting HLD smells
are depicted in Figure~\ref{fig:Hub-Prediction-Results}. 
The filter allowed us to find all the correct hubs in every triple of analyzed versions. However, this result came at the cost of increasing the number of nodes mistakenly predicted, which manifested as a drop in precision. 

From the observed results, we can say that that even though hubs
could appear not only due to the addition of new edges to the specific
nodes, but also due to changes in the overall graph structure, our analysis of the full set of predicted edges misguided the smell predictions, thus resulting in inaccurate results. 
The implications of these results are two-fold.
First, for predicting new hubs, the analysis of nodes' neighborhood seems more informative than the overall graph structure 
Only one dependency might not be sufficient to predict that a given node might become a hub. 
Second, hub predictions might not only depend on the known structure of the graph, but also on the additional packages and dependencies that are sub-sequentially added (as the result of the prediction). 


The results for Apache
Ant reinforced the observations drawn from Apache Derby. Note that the
worst prediction results were in those system versions obtaining
the worst dependency prediction results (in the first phase). 

\subsubsection*{Threats to Validity}
The study and predictions analyzed for the systems had some threats to validity. First, the characteristics of the two systems (all from the Apache ecosystem), their selected versions, as well as the particular smell instances appearing in those versions, might have influenced the performance of the classification model. Although we acknowledge that not all triples of versions are suitable for having good predictions, the relations between the selected triples and the features used in the classification model need further analysis. Second, the features might have also favored predictions for some smell types (e.g., cycles) but not for others (e.g., hubs). Thus, extending the approach to support new types of dependency-based smells requires the implementation of new filter strategies, but also might involve adjustments in the classification model. Third, we considered each package of the system as being a different module. However, this assumption might not hold in all systems, because of code organization aspects. For example, different sub-packages might belong to the same conceptual module. The criterion for identifying modules (from the code) or the granularity at which they are considered changes the graph structure, and consequently it might affect the link prediction task.

\section{Related Work} 
\label{sec:related-work}

Regarding architectural smells, several characterizations and catalogs of smells have been reported in the literature \cite{Lippert2006,Mo2015HotspotPT,conf/qosa/GarciaPEM09,Le2016}. Most of these works categorize certain smells as related to undesired dependencies. Although several academic and commercial tools provide capabilities for automated detection of smells, only a few of them support the detection of architectural smells. Cycles are a common type of smell in existing catalogs, and its detection is supported by most tools.

In \cite{Mo2015HotspotPT}, the authors formalize the definition of five  architectural smells, called hotspot patterns, including a type of package cycle. A tool called HotspotDetector, based on design structure matrices, is able to automatically detect
instances of such hotspot patterns. A qualitative analysis with an
industrial study demonstrated that the approach helps developers not
only to find the important structural problems, but also guides them
in conducting refactorings for those problems. Arcan~\cite{DBLP:conf/icsa/FontanaPRTZN17} is a static analysis tool targeted to the detection of three architectural smells, including cycles and hubs. Arcan creates a graph database containing the structural dependencies of a Java system, and then runs several detection algorithms (one per smell) on this graph. At last, there are some commercial tools for detecting architectural smells, such
as Designite\footnote{\url{http://www.designite-tools.com}}, which
identifies seven architecture smells, including cycles and other dependency-based smell. As far as we are aware of, all the previous tools have no predictive capabilities.


When it comes to SNA techniques, a number of applications to Software Engineering problems have been reported \cite{Zimmermann:2008:PDU:1368088.1368161,Bhattacharya:2012:GAP:2337223.2337273, Nguyen:2010:PVS:1853919.1853923}. For example, SNA
has been used to predict software evolution~\cite{seke2013b,Bhattacharya:2012:GAP:2337223.2337273},
the appearance of defects and bugs~\cite{Zimmermann:2008:PDU:1368088.1368161,Bhattacharya:2012:GAP:2337223.2337273}, and the existence of vulnerable components~\cite{Nguyen:2010:PVS:1853919.1853923}.
In all cases, the authors agreed that the topological analysis of
dependency graphs can reveal (or even predict) interesting properties of the software system under analysis. Another example is \cite{aryani2014predicting} that proposes an approach for extracting domain information from user manuals and predict logical coupling among software artifacts. However, with the exception of \cite{Zhou:2014:BPM:2671850.2671886}, LP techniques have not been exploited yet in Software Engineering.  The closest LP approach for the first phase of our approach is the one proposed by Zhou et al. \cite{Zhou:2014:BPM:2671850.2671886}, which used LP techniques for predicting missing dependencies in build configuration files. Unlike our approach, ML techniques were not employed. Instead, the authors applied traditional LP algorithms derived from the homophily principle, with uneven results. A custom algorithm was later developed for the problem.




\section{Conclusions and Outlook}
\label{sec:conclusions} 

In this work, we develop an approach based on LP techniques and a classification model for predicting instances of architectural smells that are likely to appear in a system. This predictive capability is the main contribution of the work. The  smells fall in the category of dependency-related smells. The predictions are informed by the current and previous versions of the system. The classification model relies on topological features of the package dependency graphs and also on context-based features from the system source code.  The classification model works in tandem with a set of filters, according to the smell types being detected. 

An initial evaluation with two types of smells in two open-source system showed a good performance for the positive class, when inferring individual dependencies, and also showed a high recall regarding the identification of the actual smells (in the next system version). This first aspect is attributed to the inclusion of content-based features in the classification model. As for the second aspect, we found evidence that the choice of the filter variant (for a given smell type) can affect both recall and precision, although we preferred good recall over precision in the analyzed cases. Furthermore, we observed that the smell predictions depended on the overall system structure (i.e., the package graph) as well as on the version history of the particular systems.


Despite the promising results, the approach has still some drawbacks and open issues. First, we need to perform a systematic study with more systems (both commercial and open-source) and with other types of dependency-based smells, in order to corroborate our findings. Second, the prediction capabilities are sensitive to the outputs of the classifier model (first phase of the approach), and are affected to a lesser extent by the subsequent filtering phase. Therefore, we still need to analyze and possibly extend the set of features used in the classification model. Along this line, we will continue investigating content-based criteria, and also consider software-specific metrics for source code entities. Some works have suggested that previous occurrences of a smell (within a given package) can increase the chance of the smell re-appearing in future versions. This information could be added as features to the classifier. There are also reported cases in which some smells are not harmful or might correspond to good design decisions (e.g., a Visitor pattern might generate a cycle among the participating elements) \cite{DBLP:conf/wcre/FontanaDWYZ16}. We would like to analyze if these examples can be learned by the classifier and distinguished in the predictions.

As well as new dependencies are added, existing dependencies could disappear in future versions, providing another opportunity for prediction. In this case, we are interested in predicting whether the previous graph structure (i.e. when a potentially disappearing dependency did not exist) will occur again \cite{Nagrecha:2015:RSP:2808797.2809283}. 
Finally, the integration of the approach with existing tools or research prototypes, such as SonarQube or Arcan \cite{DBLP:conf/icsa/FontanaPRTZN17}, is another subject for future work.
\bigskip



\footnotesize
\bibliographystyle{plainnat}

\end{document}